\documentclass[11pt]{article}
\usepackage{a4wide,graphicx,amsmath,cite,longtable,multicol,array}

\makeatletter
\@addtoreset{equation}{section}

\makeatother

\newcommand\bsg{b\to s\gamma}
\newcommand\Hbs{H\to b s}
\newcommand\Hzbs{H^0\to b s}
\newcommand\ii{\mathrm{i}}
\newcommand\Ord{\mathcal{O}}
\newcommand\BR{\mathcal{B}}
\newcommand\LL{\mathrm{LL}}
\newcommand\LR{\mathrm{LR}}
\newcommand\RL{\mathrm{RL}}
\newcommand\RR{\mathrm{RR}}
\newcommand\de[1]{(\delta_{#1})_{23}}
\newcommand\deu[1]{(\delta_{#1}^u)_{23}}
\newcommand\ded[1]{(\delta_{#1}^d)_{23}}
\newcommand\deud[1]{(\delta_{#1}^{u,d})_{23}}
\newcommand\MQ[1]{M_{\tilde Q,#1}}
\newcommand\MU[1]{M_{\tilde U,#1}}
\newcommand\MD[1]{M_{\tilde D,#1}}
\newcommand\ML[1]{M_{\tilde L,#1}}
\newcommand\MR[1]{M_{\tilde R,#1}}
\newcommand\MGl{M_{\tilde g}}
\newcommand\diag{\mathop{\mathrm{diag}}}
\newcommand\sub[1]{_{\text{#1}}}

\newenvironment{CoupVec}%
  {\left[\begin{array}{>{\displaystyle}c}}%
  {\end{array}\right]}
\newenvironment{PlusB}%
  {\left\{\begin{array}{l}}%
  {\end{array}\right\}}

\newcommand\Coupling[2]{\par%
  {\large\texttt{[#1]}~~\textbf{#2}}\\[3ex]%
  \nopagebreak\bigskip\ignorespaces%
}

\begin{document}

\begin{titlepage}

\hfill\begin{tabular}{l}
MPP-2005-168 \\
CAFPE-67/05 \\
UG-FT-197/05\\
CERN-PH-TH/2005-208
\end{tabular}

\bigskip
\bigskip
\bigskip
\bigskip

\begin{center}

\begin{LARGE}
\textbf{\boldmath{Interplay between $H\to b\bar s$ and $\bsg$ \\[1ex]
in the MSSM with Non-Minimal Flavour Violation}}
\end{LARGE}

\bigskip
\bigskip
\bigskip
\bigskip

{\Large T.~Hahn$^a$, W.~Hollik$^a$, J.I.~Illana$^b$,    S.~Pe{\~n}aranda$^c$}~\footnote{%
        e-mails: hahn@mppmu.mpg.de, hollik@mppmu.mpg.de,
        jillana@ugr.es, siannah.penaranda@cern.ch}

\bigskip
\bigskip

\begin{it}
$^a$Max-Planck-Institut f\"ur Physik, F\"ohringer Ring 6,
D--80805 Munich, Germany

\smallskip

$^b$CAFPE and Departamento de F{\'\i}sica Te\'orica y del
Cosmos, Universidad de Granada, E--18071 Granada, Spain

\smallskip

$^c$CERN TH Division, Department of Physics,
CH--1211 Geneva 23, Switzerland
\end{it}

\bigskip
\bigskip
\bigskip
\bigskip

\textbf{Abstract}

\end{center}

\noindent
We investigate the constraints on flavour-changing neutral heavy
Higgs-boson decays $H\to b\bar s$ from $\bsg$ bounds on the
flavour-mixing parameters of the MSSM with non-minimal flavour violation
(NMFV).  In our analysis we include the contributions from the SM and
new physics due to general flavour mixing in the squark mass matrices.
We study the case of one and two non-zero flavour-mixing parameters and
find that in the latter case the interference can raise the Higgs
flavour-changing branching ratios by one or two orders 
of magnitude with respect
to previous predictions based on a single non-zero parameter and in
agreement with present constraints from $B$ physics.  In the course of
our work we developed a new \textsl{FeynArts} model file for the NMFV
MSSM and added the necessary code for the evaluation to
\textsl{FormCalc}.  Both extensions are publicly available.

\end{titlepage}


\section{Introduction}

Flavour-changing neutral current (FCNC) processes provide an
extraordinarily useful tool to investigate new physics beyond the
Standard Model (SM). In supersymmetry (SUSY), as in other models beyond
the SM, an alternative to the direct search for new particles is to look
for their radiative effects.  In the SM, FCNCs are absent at tree level,
but they arise at one-loop order, being strongly suppressed as a consequence
of the GIM mechanism~\cite{GIM}.  The Minimal
Supersymmetric Standard Model (MSSM)~\cite{HaberKane} provides a natural
framework where FCNCs are enhanced.  In the scenario with minimal
flavour violation (MFV) of the MSSM, squarks are assumed to be aligned
with the corresponding quarks, and flavour violation originates from the
Cabibbo--Kobayashi--Maskawa (CKM) matrix as the only source, proceeding
via loop contributions, in analogy to the SM.  Therefore, its size is
expected to be very small.  In more general scenarios which include
misalignment between the quark and squark sectors, sizeable
contributions to FCNC processes are expected to occur.  This is the case
of the MSSM with non-minimal flavour violation (NMFV).

The quantum contributions from supersymmetric particles in $B$-meson
physics have been studied in a series of processes, including
$B^0$--$\bar B^0$ mixing~\cite{toharia,newchanko}, leptonic $B$-meson
decays~\cite{kolda,oldchanko,alemanes,newisidori} and $B\to
X_s\gamma$~\cite{bstanbeta,bsgamma,leadingbsg,gambinobsg,hurth}, and
have been found to be large.  In order not to be in conflict with
present experimental data, these in turn imply restrictions both on the
SUSY parameters and on the parameters measuring the size of flavour
mixing in the squark sector~\cite{newisidori,hurth,9604387,ciuchini}. 
The rare decay $B\to X_s \gamma$ is at present one of the most important
since its observation~\cite{boundsbsg} sets stringent constraints on the
parameter space of various extensions of the
SM~\cite{gambinobsg,9604387,hurth}. 
Furthermore, the recent data from $B\to X_s\mu^+\mu^-$ indicate that the 
sign of the $\bsg$ amplitude is the same as in the SM \cite{Gambino:2004mv}.

Moreover, it is well known that FCNC processes related to Higgs-boson
physics are also very sensitive to supersymmetric quantum
effects~\cite{Maria,nos,dedes,Demir,radFCNC,santi,kolda,oldchanko,alemanes,newisidori}.
In particular, large rates of neutral Higgs decays into two quarks of
different flavours have been predicted~\cite{Maria,nos,santi}.\footnote{%
        In the following, $\BR(\Hbs)$ denotes the sum of the Higgs 
        branching ratios into $b\bar s$ and $\bar bs$.  The Higgs boson 
        $H$ stands for that of the SM, $H\sub{SM}$, or one of 
        those of the MSSM, $h^0$, $H^0$, or $A^0$.}
In the SM, one finds $\BR(H\sub{SM}\to bs) \approx 4\times 10^{-8}$ for
$m_{H\sub{SM}}= 114$ GeV.  For the neutral MSSM Higgs bosons the 
ratios could be of $\Ord(10^{-4}$--$10^{-3})$.  Constraints from $\bsg$
data reduce these rates, though~\cite{santi,nos}.

In this paper we provide a phenomenological analysis of the general
constraints on flavour-changing neutral Higgs decays $H\to b\bar s,
s\bar b$, set by bounds from $\bsg$ on the flavour-mixing parameters in
the squark mass matrices of the MSSM with NMFV.  We include the SM and
the full genuine SUSY contributions by taking into account their
interference and, in particular, the influence of several
flavour-changing parameters contributing simultaneously.  Since the full
diagrammatic approach is used, our computation is valid for all values
of the characteristic parameter measuring the squark-mixing strength,
beyond the mass-insertion approximation, and for all values of
$\tan\beta$.  Previous analyses of bounds on SUSY flavour-mixing
parameters from $\bsg$~\cite{hurth} have shown the importance of the
interference effects between the different types of flavour
violation~\cite{9604387}.  In the present work we derive predictions for
$\BR(\Hbs)$ compatible with present experimental $\bsg$ bounds 
and recent data from $B\to X_s\mu^+\mu^-$, assuming
first one and then several types of flavour mixing contributing at a
time for comparison.

The paper is organized as follows.  In Section \ref{sect:FC_MSSM} the
squark mass matrices in the MSSM with NMFV are described in detail and
the notation is introduced.  The numerical analysis of the branching
ratios $\BR(\Hbs)$ 
compatible with the observed decay rates for $B\to X_s\gamma$ and
$B\to X_s\mu^+\mu^-$ is included in
Section \ref{sect:Higgsdecays}.  In that section, we consider only one
specific flavour-mixing parameter different from zero and discuss the
interference effects of the SM and the new physics contributions.  The
interplay between $\Hbs$ and $\bsg$ is presented in Section
\ref{sect:interplay}.  The constraints on flavour-changing neutral heavy
Higgs-boson decays due to different types of flavour violation are
derived there.  Conclusions are given in Section 5.  The relevant
Feynman rules are listed in Appendix \ref{sect:frules}.


\section{Non-minimal Flavour Mixing in the MSSM}
\label{sect:FC_MSSM}

In the MSSM there are two sources of flavour violation.  The first one
arises from different rotations of the $d$- and $u$-quark fields
from the interaction to the physical bases, and its strength is driven
by the off-diagonal CKM-matrix elements, as in the SM.  
The second one consists of a
possible misalignment between the rotations that
diagonalize the quark and squark sectors (NMFV).  
The part of the soft-SUSY-breaking Lagrangian responsible for 
this non-minimal squark family mixing is given by
\begin{align}
\label{eq:lagrangian}
\mathcal{L}^{\text{squark}}_{\text{soft}} &=
-\tilde Q_i^\dagger (M_{\tilde Q}^2)_{ij} \tilde Q_j
-\tilde U_i^\dagger (M_{\tilde U}^2)_{ij} \tilde U_j
-\tilde D_i^\dagger (M_{\tilde D}^2)_{ij} \tilde D_j \nonumber \\
&\quad + y^u_i A^u_{ij} \tilde Q_i H_u \tilde U_j
+ y^d_i A^d_{ij} \tilde Q_i H_d \tilde D_j\,,
\end{align}
where $\tilde Q$ is the SU(2) scalar doublet, $\tilde U$, $\tilde D$ are
the up- and down-squark SU(2) singlets, respectively, $y^{u,d}$ are the
Yukawa couplings and $i,j$ are generation indices.  The flavour-changing
effects come from the non-diagonal entries in the bilinear terms
$M_{\tilde Q}^2$, $M_{\tilde U}^2$, and $M_{\tilde D}^2$, and from the 
trilinear terms $A_u$ and $A_d$.

We assume that the non-CKM squark mixing is significant only for 
transitions between the
squarks of the second and third generations.  They are expected to be
the largest in Grand Unified Models and are also experimentally the
least constrained.  The most stringent bounds are set by $\bsg$.  In
contrast, there exist strong experimental bounds involving the first
squark generation, based on data from $K^0$--$\bar K^0$ and $D^0$--$\bar
D^0$ mixing~\cite{9604387,ciuchini}.

Our parameterization of the flavour-non-diagonal squark mass
matrices for the up- and down-type squarks, for the MSSM with real
parameters, reads as follows,
\begin{equation}
\label{eq:usquarkmass}
M^2_{\tilde u} = \left(\begin{array}{ccc|ccc}
\ML{u}^2 & 0 & 0 & m_u X_u & 0 & 0 \\
0 & \ML{c}^2 & \Delta_\LL^u & 0 & m_c X_c & \Delta_\LR^u \\
0 & \Delta_\LL^u & \ML{t}^2 & 0 & \Delta_\RL^u & m_t X_t \\[.3ex]
\hline
m_u X_u & 0 & 0 & \MR{u}^2 & 0 & 0 \\
0 & m_c X_c & \Delta_\RL^u & 0 & \MR{c}^2 & \Delta_\RR^u \\
0 & \Delta_\LR^u & m_t X_t & 0 & \Delta_\RR^u & \MR{t}^2
\end{array}\right)\,,
\end{equation}
\begin{equation}
\label{eq:dsquarkmass}
M^2_{\tilde d} = \left(\begin{array}{ccc|ccc}
\ML{d}^2 & 0 & 0 & m_d X_d & 0 & 0 \\
0 & \ML{s}^2 & \Delta_\LL^d & 0 & m_s X_s & \Delta_\LR^d \\
0 & \Delta_\LL^d & \ML{b}^2 & 0 & \Delta_\RL^d & m_b X_b \\[.3ex]
\hline
m_d X_d & 0 & 0 & \MR{d}^2 & 0 & 0 \\
0 & m_s X_s & \Delta_\RL^d & 0 & \MR{s}^2 & \Delta_\RR^d \\
0 & \Delta_\LR^d & m_b X_b & 0 & \Delta_\RR^d & \MR{b}^2
\end{array}\right)\,,
\end{equation}
where
\begin{equation}
\begin{aligned}
\label{eq:squarkparam}
\ML{q}^2 &=
  M_{\tilde Q,q}^2 + m_q^2 + \cos2\beta (T_3^q - Q_q s_W^2) m_Z^2\,, \\
\MR{\{u,c,t\}}^2 &=
  M_{\tilde U,\{u,c,t\}}^2 + m_{u,c,t}^2 + \cos2\beta Q_t s_W^2 m_Z^2\,, \\
\MR{\{d,s,b\}}^2 &=
  M_{\tilde D,\{d,s,b\}}^2 + m_{d,s,b}^2 + \cos2\beta Q_b s_W^2 m_Z^2\,, \\
X_{u,c,t} &= A_{u,c,t} - \mu\cot\beta\,, \\
X_{d,s,b} &= A_{d,s,b} - \mu\tan\beta\,,
\end{aligned}
\end{equation}
with $m_q$, $T_3^q$, $Q_q$ the mass, isospin, and electric charge of the
quark $q$, $m_Z$ the $Z$-boson mass, $s_W \equiv \sin\theta_W$,
$\theta_W$ the electroweak mixing angle, and $\mu$ the Higgsino mass
parameter. 

We define the dimensionless flavour-changing parameters
$(\delta^{u,d}_{ab})_{23}$ $(ab = \LL,\LR,\RL,\RR)$ from the 
flavour-off-diagonal elements of the squark mass matrices 
(\ref{eq:usquarkmass}) and (\ref{eq:dsquarkmass}) in the following way,
\begin{equation}
\begin{aligned}
\label{eq:FCparam}
\Delta_\LL^u &\equiv \deu\LL \ML{c} \ML{t}\,,\quad &
\Delta_\LL^d &\equiv \ded\LL \ML{s} \ML{b}\,, \\
\Delta_\LR^u &\equiv \deu\LR \ML{c} \MR{t}\,,\quad &
\Delta_\LR^d &\equiv \ded\LR \ML{s} \MR{b}\,, \\
\Delta_\RL^u &\equiv \deu\RL \MR{c} \ML{t}\,,\quad &
\Delta_\RL^d &\equiv \ded\RL \MR{s} \ML{b}\,, \\
\Delta_\RR^u &\equiv \deu\RR \MR{c} \MR{t}\,,\quad &
\Delta_\RR^d &\equiv \ded\RR \MR{s} \MR{b}\,.
\end{aligned}
\end{equation}
In our phenomenological study, they are free parameters determining the
size of NMFV induced by SUSY and are analogous to those defined in the
mass-insertion approximation~\cite{9604387}.

In order to diagonalize the two $6\times 6$ squark-mass matrices given
above, the $6\times 6$ matrices $R^{\tilde u}$ for the up-type squarks and 
$R^{\tilde d}$ for the down-type squarks are needed,
\begin{equation}
\tilde u_\sigma = R^{\tilde u}_{\sigma,j} \begin{pmatrix}
  \tilde u_L \\
  \tilde c_L \\
  \tilde t_L \\
  \tilde u_R \\
  \tilde c_R \\
  \tilde t_R
\end{pmatrix}_j\,, \quad
\tilde d_\sigma = R^{\tilde d}_{\sigma,j} \begin{pmatrix}
  \tilde d_L \\
  \tilde s_L \\
  \tilde b_L \\
  \tilde d_R \\
  \tilde s_R \\
  \tilde b_R
\end{pmatrix}_j\,.
\end{equation}

The diagonalization yields the squark mass eigenvalues and eigenstates
depending on the flavour-mixing parameters $\de{ab}$, i.e.\ $R^{\tilde
q} M_{\tilde q}^2 R^{\tilde q\dagger} = \diag(M^2_{\tilde
q_1},\dots,M^2_{\tilde q_6})$.  The dependence of the squark masses on
$\de\LL$ has already been studied in~\cite{Maria}.  Typically, out of
the four eigenvalues involving the second and third generations, two are
weakly dependent on the amount of flavour mixing and for the other two,
one grows and the other decreases with $\de\LL$.  In general, 
flavour mixing through the flavour non-diagonal entries in the squark
mass matrices generates large splittings between the squark mass
eigenvalues.


\section{Flavour-changing decay processes}
\label{sect:Higgsdecays}

We focus here on the loop-induced flavour-changing decays of the MSSM
heavy neutral Higgs bosons $H = H^0, A^0$ into second- and
third-generation quarks, $H\to b\bar s, s\bar b$
(an independent analysis for the lightest Higgs boson $h^0$
will be reported elsewhere~\cite{future}).
We include the
contributions from SM particles and their superpartners (squarks,
gluinos, charginos, and neutralinos), as well as those from the MSSM
Higgs sector, and also their interference effects.  We use the full
diagrammatic approach for arbitrary values of $\tan\beta$  and of the
flavour-mixing parameters $\deud{ab}$.  
Note that $\bsg$ constrains
$\ded{ab}$ only.  For simplicity, we take the same values for the
flavour-mixing parameters in the up- and down-squark sectors:
$\de{ab}\equiv\deu{ab} = \ded{ab}$.  Actually, the LL blocks of the up-
and down-squark mass matrices are related by the SU(2)$_L$ gauge
symmetry \cite{9604387}, therefore a large difference between $\deu\LL$
and $\ded\LL$ is not allowed.  For the same reason,
$\MQ{u}\approx\MQ{d}$, $\MQ{c}\approx\MQ{s}$, and $\MQ{t}\approx\MQ{b}$.

We have taken the expression for the branching ratio $\BR(B\to X_s\gamma)$ to NLO
from \cite{Kagan}. In the MSSM with NMFV, the relevant operators of the effective 
Hamiltonian are
\begin{equation}
\begin{aligned}
O_2 &= \bar s_L\gamma_\mu c_L\gamma^\mu b_L ,\\
O_7 &= \frac{e}{16\pi^2}m_b \bar s_L \sigma_{\mu\nu}F^{\mu\nu} b_R,\quad
&\tilde O_7 &= \frac{e}{16\pi^2}m_b \bar s_R \sigma_{\mu\nu}F^{\mu\nu} b_L,\\ 
O_8 &= \frac{g_s}{16\pi^2}m_b \bar s_L\sigma_{\mu\nu}G^{\mu\nu}_a t_a b_R,\quad
&\tilde O_8 &= \frac{g_s}{16\pi^2}m_b \bar s_R\sigma_{\mu\nu}G^{\mu\nu}_a t_a b_L.\\
\end{aligned}
\end{equation}
We have calculated the corresponding Wilson coefficients $C_{2,7,8}$ and
$\tilde C_{7,8}$ to one loop. The tilded operators do not contribute in the SM or 
in the MSSM with MFV, in the limit of massless strange quark. 
The data from $B\to X_s\mu^+\mu^-$ require that the sign of the coefficient 
$C_7(m_b)$ is the same as in the SM.

We used \textsl{FeynArts}, \textsl{FormCalc}, and \textsl{LoopTools} to
obtain our results.  To this end, we had to modify the MSSM model file of
\textsl{FeynArts} to include general flavour mixing,  and  we added
$6\times 6$ squark mass and mixing matrices to the \textsl{FormCalc}
evaluation \cite{hahn,thtalk}.  A list of the Feynman rules needed for our
computation is given in Appendix A.
The masses and total decay widths of the Higgs bosons were computed with 
{\sl FeynHiggs}~\cite{Feynhiggs}.

For a concrete evaluation, 
we choose the following six flavour-diagonal MSSM parameters
as input parameters: $m_A$,
$\tan \beta$, $\mu$, $M_{2}$, $M\sub{SUSY}$, $A$, where $m_A$ denotes the mass
of the CP-odd Higgs boson $A^0$.
For simplicity, we have taken a common value for the soft SUSY-breaking
squark mass parameters $M\sub{SUSY}\equiv \MU{\{u,t,c\}} =
\MD{\{d,b,s\}}$, and all the various trilinear parameters to be
universal, $A\equiv A_t = A_b = A_c = A_s$.  These parameters and the
$\delta$'s will be varied over a wide range, subject only to the
requirements that all the squark masses be heavier than 100 GeV, 
$|\mu| > 90$ GeV and $M_2 > 46$ GeV~\cite{pdg2004}.  

The following MSSM parameters have been chosen as a default set 
(if not specified differently),
\begin{equation}
\label{eq:numparameters}
\begin{gathered}
M\sub{SUSY} = 800\text{ GeV}\,, \quad
M_2 = 300\text{ GeV}\,, \quad
M_1 = \frac 53 \frac{s_W^2}{c_W^2} M_2\,, \\
A   = 500\text{ GeV}\,, \quad
m_A = 400\text{ GeV}\,, \quad
\tan\beta = 35\,, \quad
\mu = -700\text{ GeV}\,.
\end{gathered}
\end{equation}
A detailed study of the values of the Higgs-boson decay widths in
a wider range of these parameters and for $\deu\LL \neq 0$ 
has been done in~\cite{nos}.  

Let us remark that the $\mu$-parameter plays an important role in our
analysis.  The $H^0$ decay width is approximately symmetric under $\mu
\to -\mu$, depending on the $m_A$ values, and increases with $\mu$ up to
a certain value around $600$ GeV; then it reaches a maximum value, and
finally decreases~\cite{Maria,nos}.  Moreover, the $\mu$-parameter
enters in the gluino and chargino contributions to $\bsg$, producing
significant cancellations at large $\tan\beta$. 

For illustration, we give in this section an overview over 
the various contributions
to both $\Hzbs$ and $B \to X_s\gamma$, keeping only a single
flavour-off-diagonal element, $\de\LL$. For the Higgs decays, we agree with 
the results of previous studies on the subject~\cite{Maria,nos,santi}.

\begin{figure}
\begin{center}
\begin{tabular}{cc}
\includegraphics[width=.45\hsize]{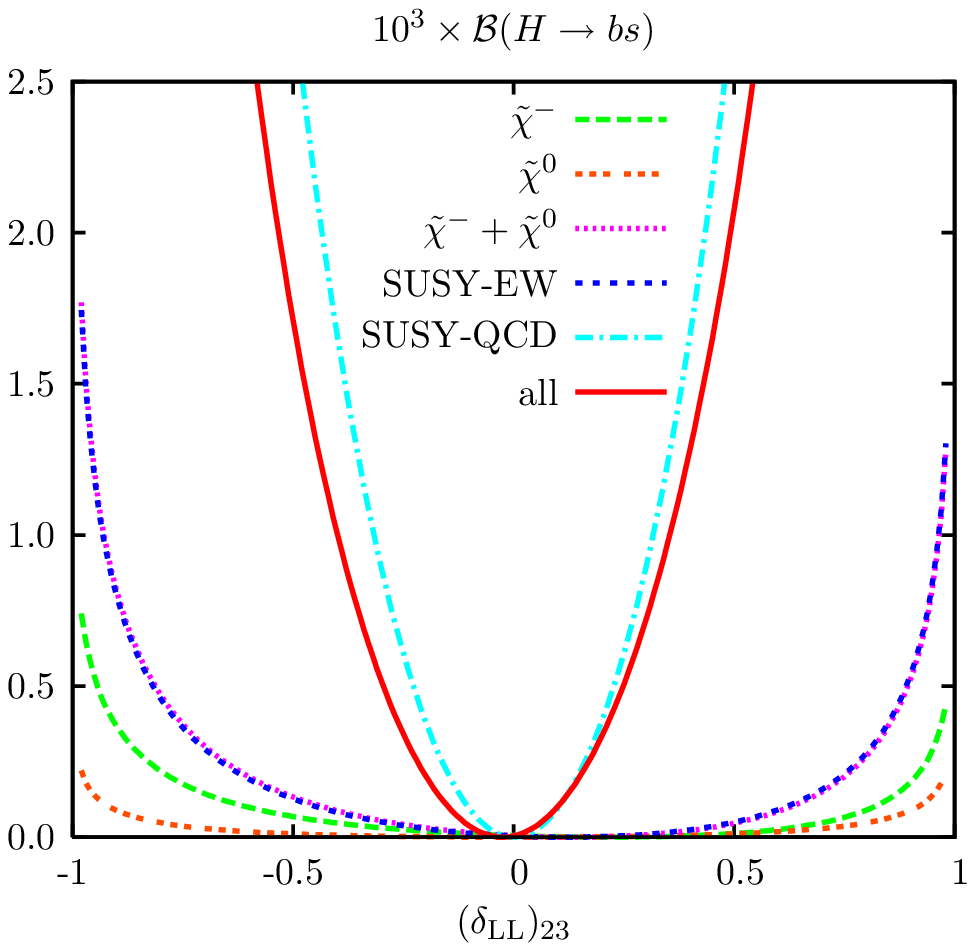} &
\includegraphics[width=.45\hsize]{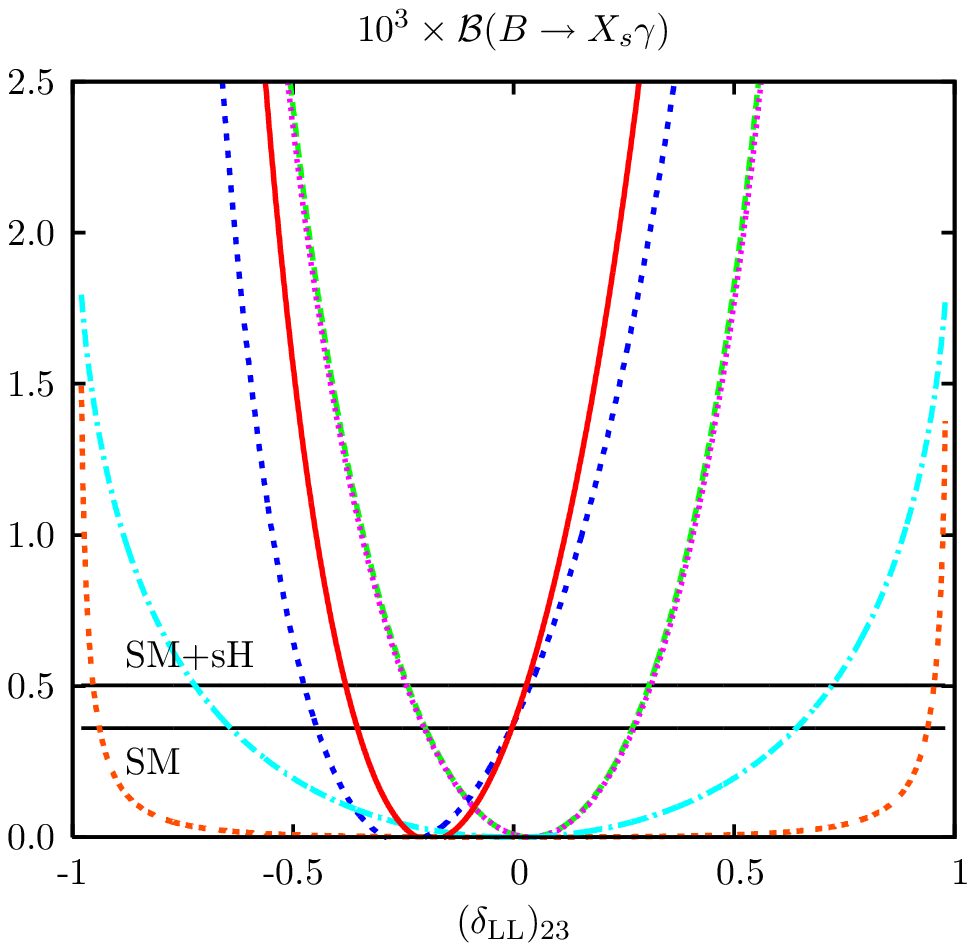} \\
\includegraphics[width=.45\hsize]{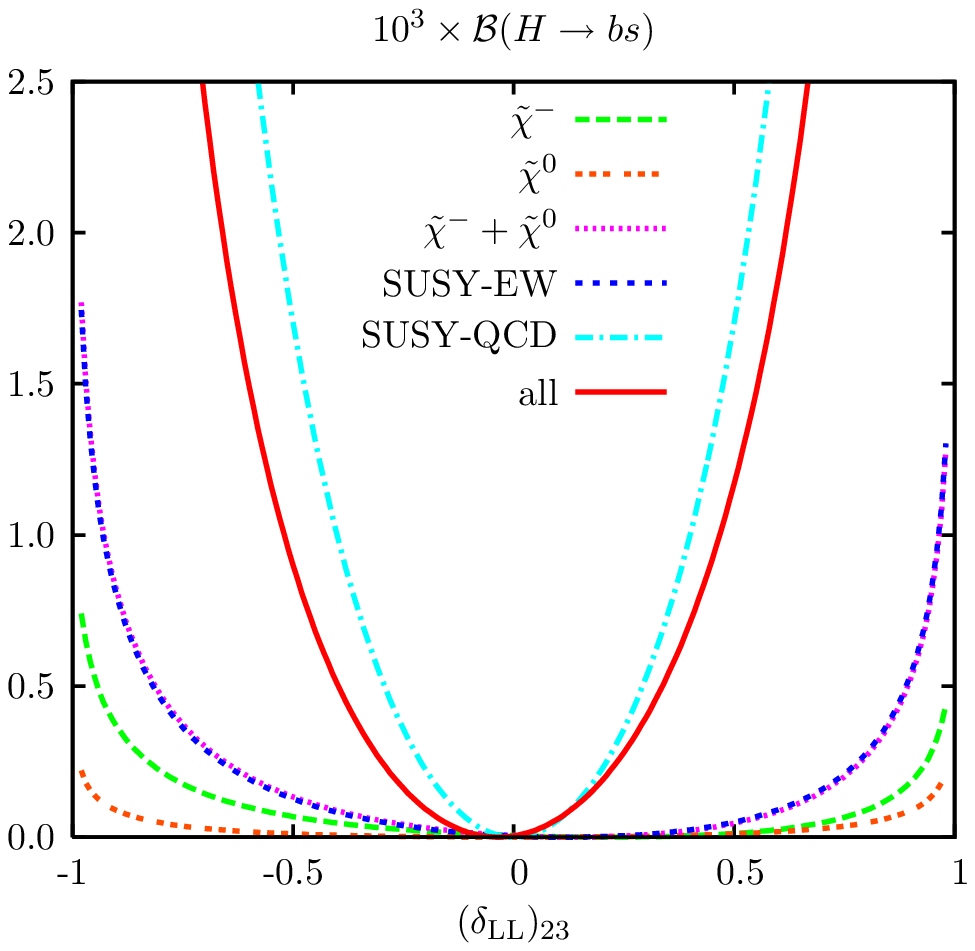} &
\includegraphics[width=.45\hsize]{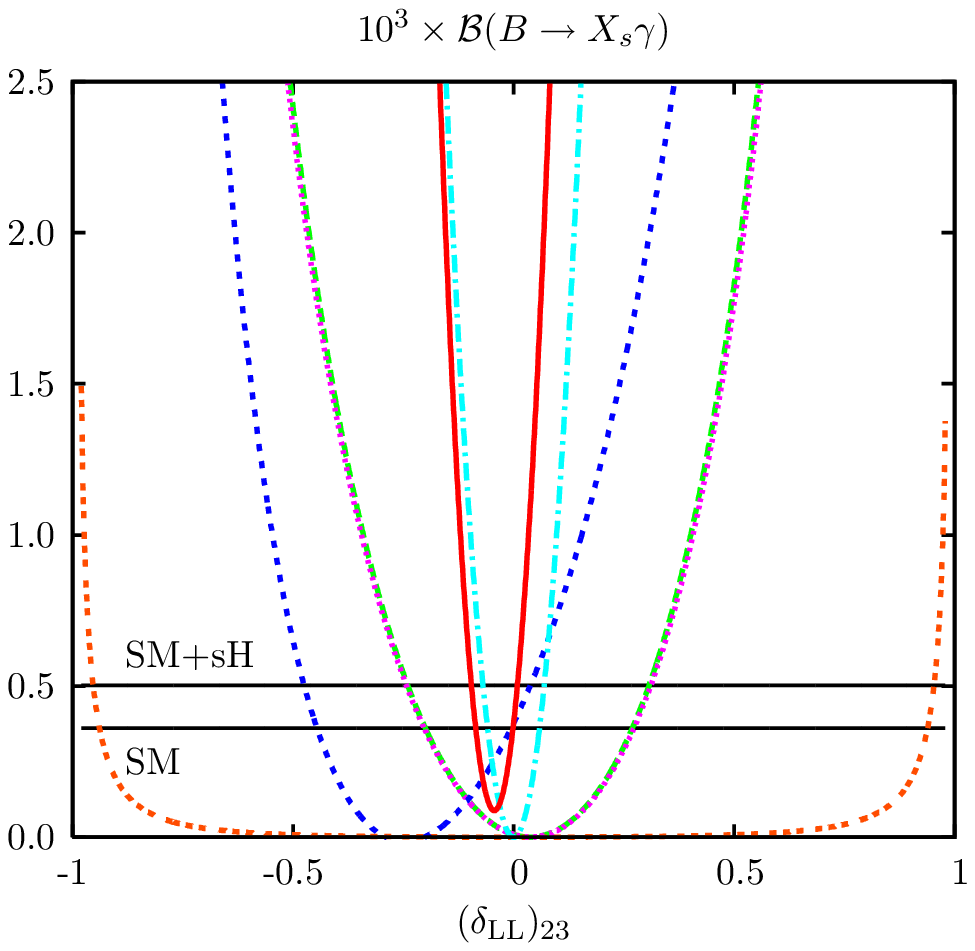} \\
\end{tabular}
\end{center}
\vspace*{-3ex}
\caption{\label{fig:contrib}$\BR(\Hzbs)$ [left] and
$\BR(B\to X_s\gamma)$ [right] as a function of $\de\LL$.  In the
upper row $\MGl$ is determined from $M_2$ through the GUT relation, in
the lower row $\MGl = 300$~GeV is chosen.  The other input parameters
are given in Eq.~(\ref{eq:numparameters}). The SM and the MSSM Higgs 
sector contributions (sH) are included (horizontal lines in 
$\BR(B\to X_s\gamma)$ [right], but invisible in $\BR(\Hzbs)$ [left]).}
\end{figure}

Fig.~\ref{fig:contrib} shows on the left $\BR(\Hzbs)$ and on the right
$\BR(B\to X_s\gamma)$ as a function of $\de\LL$.  In the upper row the GUT
relation $M_3 = \alpha_s/(\alpha s_W^2) M_2$ is assumed, yielding the
rather large gluino mass of $\MGl\approx 1$ TeV. In the lower row $\MGl
= 300$ GeV is chosen.

Our purpose is to show the size of the radiative corrections separately
for the different sectors and their interference effects.  The SUSY
electroweak (SUSY-EW) contribution includes charginos, neutralinos, and
the MSSM Higgs sector, the latter being negligible in $\Hzbs$ but
important in $\bsg$ (horizontal lines in $\BR(\bsg)$ because they do not 
depend on $\de\LL$). The chargino and neutralino contributions are
plotted independently to show that in both cases the charginos dominate.
Gluino--squark loops constitute the SUSY-QCD contribution. 

$\BR(\Hzbs)$ increases with $\de\LL$, as already 
known~\cite{Maria,nos,santi}.  
We can see that the SUSY-QCD contribution is dominant,
at least one order of magnitude larger than that of SUSY-EW.  Both
interfere with opposite signs, as discussed in~\cite{nos}.  The SM value
for the Higgs branching ratio is several orders of magnitude smaller
than the corresponding MSSM value, of $\Ord(10^{-8})$.

In $\bsg$, the SUSY-QCD contribution is also dominant in most of the
flavour parameter space, but becomes subdominant for large gluino masses
(see the upper row in Fig.~\ref{fig:contrib}).  
This result, however, strongly depends on the choice of parameters 
and changes if the GUT relation is relaxed.
The interference effects of the
various MSSM sectors in $\bsg$ must therefore be carefully considered.

\section{Compatibility between $\Hzbs$  and $\bsg$}
\label{sect:interplay}

Next we derive the maximum values of $\BR(H^0 \to b s)$ compatible with
$\BR(B\to X_s\gamma)\sub{exp} = (3.3\pm 0.4)\times 10^{-4}$
\cite{boundsbsg} within three standard deviations by varying the
flavour-changing parameters of the squark mass matrices.
The results for the $A^0$ boson are very similar and we do not show them 
separately.

\subsection{One flavour-mixing parameter}

As a first step, we select one possible type of flavour violation in the
squark sector, assuming that all the others vanish.  The interference
between different types of flavour mixing is thus ignored.  Previous
analyses on bounds coming from $\bsg$, by using the mass-insertion
approximation and by neglecting any kind of interference effects, led to
bounds on the single off-diagonal element for the down sector of
$|\ded{\LL,\RR}| < \Ord(1)$, $|\ded{\LR,\RL}| < \Ord(10^{-2})$
(see~\cite{9604387,hurth} and references therein).  The MSSM inputs are
those in Eq.~(\ref{eq:numparameters}) with GUT relations.

\begin{figure}
\begin{center}
\includegraphics{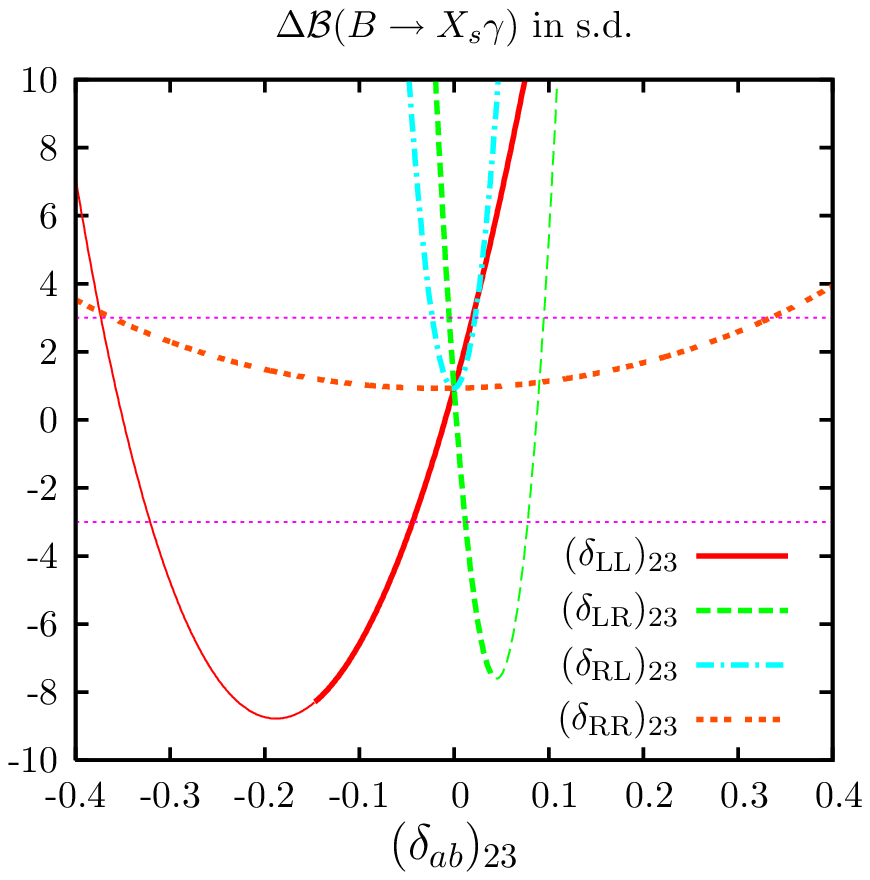}
\end{center}
\vspace*{-3ex}
\caption{\label{fig:stddev}$\Delta\BR(B\to X_s\gamma)$ in standard 
deviations (see text) as a function of $\de{\LL,\LR,\RL,\RR}$. The thinner
lines for $\de{\LL}$ and $\de{\LR}$ correspond to regions disfavoured by
$B\to X_s\mu^+\mu^-$.}
\end{figure}

In Fig.~\ref{fig:stddev} we show $\BR(B\to X_s\gamma)$ as a function of the
flavour parameters $\de{ab}$ with $ab = \LL,\LR,\RL,\RR$.  The results 
are expressed in standard deviations (s.d.) through
\begin{equation}
\frac{\Delta\BR(B\to X_s\gamma)}{1\text{ s.d.}}=
  \frac{\BR(B\to X_s\gamma)-\BR(B\to X_s\gamma)\sub{exp}}{\Delta\BR\sub{exp}}\,.
\label{DeltaBR}
\end{equation}

We can see in Fig.~\ref{fig:stddev} that the flavour-off-diagonal
elements are independently constrained to be at most $\de{ab}\sim
10^{-3}$--$10^{-1}$.  As expected~\cite{9604387,hurth}, the bounds on
$\de\LR$ are the strongest, $\de\LR\sim 10^{-3}$--$10^{-2}$. The
data from $B\to X_s\mu^+\mu^-$ further constrain the parameters $\de{\LL}$
and $\de{\LR}$, the others remaining untouched.

\begin{figure}
\begin{center}
\begin{tabular}{cc}
\includegraphics[width=.45\hsize]{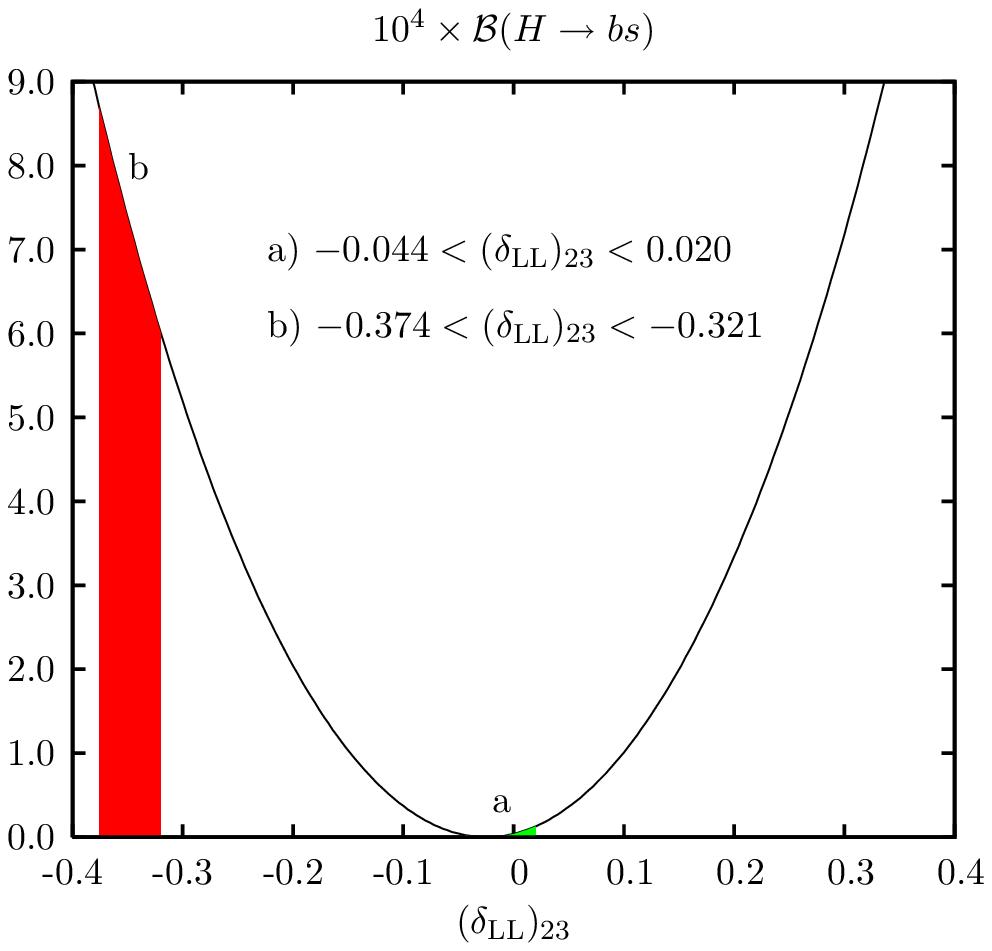} &
\includegraphics[width=.45\hsize]{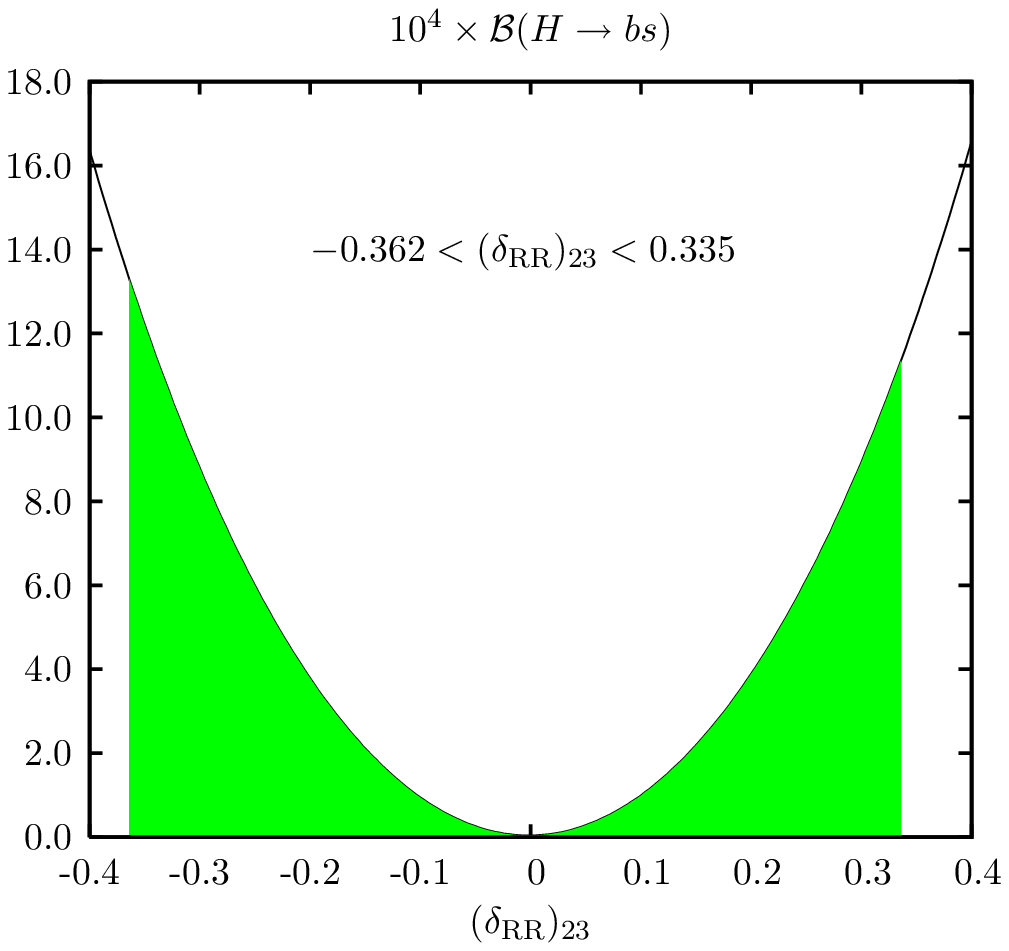} \\
\includegraphics[width=.45\hsize]{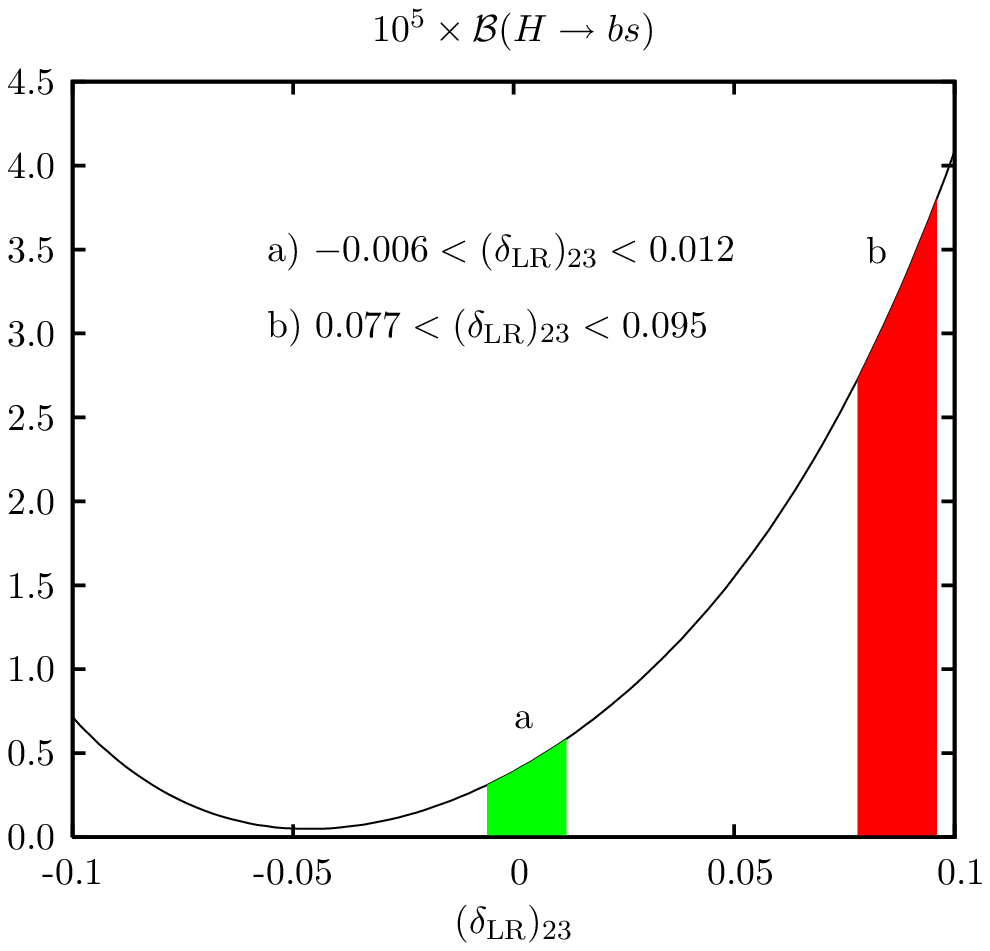} &
\includegraphics[width=.45\hsize]{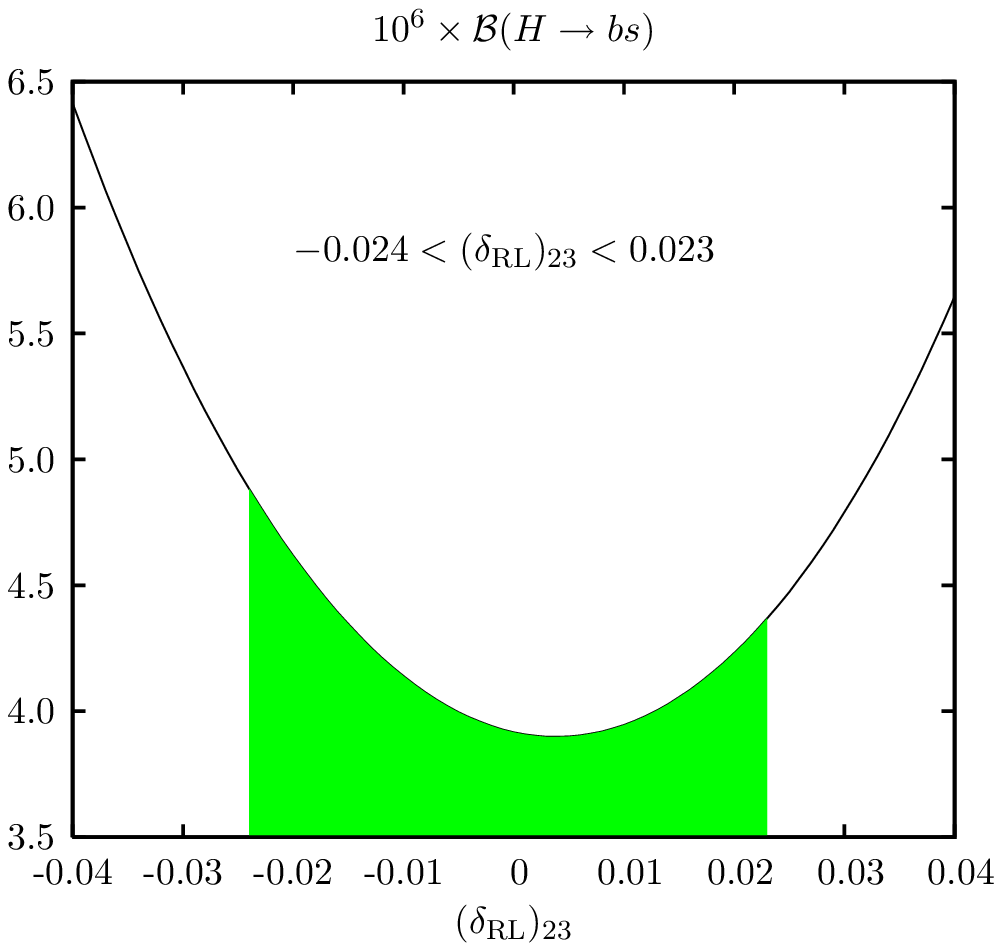}
\end{tabular}
\end{center}
\vspace*{-3ex}
\caption{\label{fig:Hbs}$\BR(\Hzbs)$ as a function of
$\de{\LL,\RR,\LR,\RL}$.  The allowed intervals of these 
parameters determined from $\bsg$ (see Fig.~\ref{fig:stddev}) are
indicated by coloured areas. The red areas (b) are disfavoured by
$B\to X_s\mu^+\mu^-$.}
\end{figure}

The allowed intervals for the corresponding flavour-mixing parameters
thus obtained are indicated in
Fig.~\ref{fig:Hbs}, in which we present the results for $\BR(\Hzbs)$ as
a function of $\de{ab}$.  The predictions compatible with the $\bsg$
constraints can be read off directly from there.

For our reference point (\ref{eq:numparameters}) we find that the
largest allowed value of $\BR(\Hzbs)$, of $\Ord(10^{-3})$ or $\Ord(10^{-5})$, 
is induced by $\de\RR$ or $\de\LL$, respectively.
These are the flavour-changing parameters least
stringently constrained by the $\bsg$ data.  $\BR(\Hzbs)$ can reach
$\Ord(10^{-6})$ if induced by $\de\LR$ or by $\de\RL$, the most stringently 
constrained flavour-changing parameter.
We remark that, because of the restrictions imposed by $\bsg$, $\BR(\Hzbs)$
depends very little on $\de\LR$ and $\de\RL$. 

\subsection{Two flavour-mixing parameters}

In this second part of our analysis, we investigate whether the maximum
values reachable by the $H^0$ branching ratios remain stable when
several off-diagonal elements of the squark mass matrix contribute
simultaneously.  It is known that the participation of several types of
flavour-changing parameters weaken the bounds, imposed by $\bsg$, on the
off-diagonal elements of the squark-mass matrix by at least one order of
magnitude~\cite{hurth}.  We derive bounds on these flavour-mixing
parameters by switching on simultaneously two of those parameters, with
all the others vanishing.  Indeed, we performed the analysis for all
possible combinations of two of the four dimensionless parameters
(\ref{eq:FCparam}).

\begin{figure}
\begin{center}
\begin{tabular}{cc}
\includegraphics[height=.42\hsize]{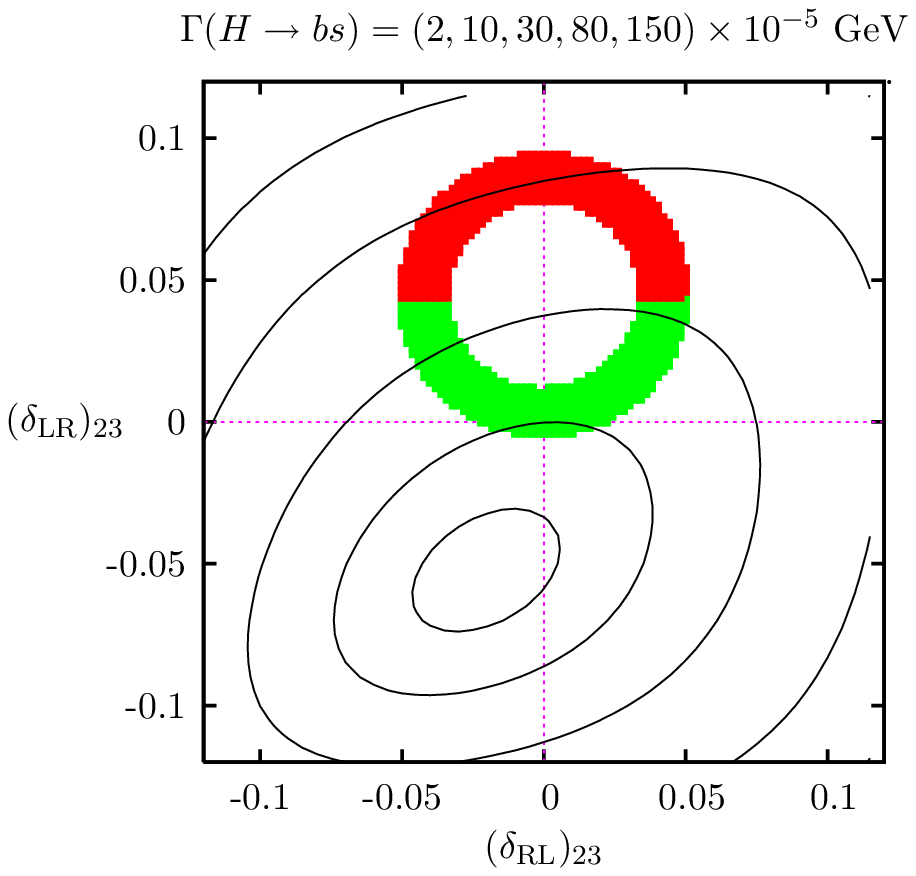} &
\includegraphics[height=.42\hsize]{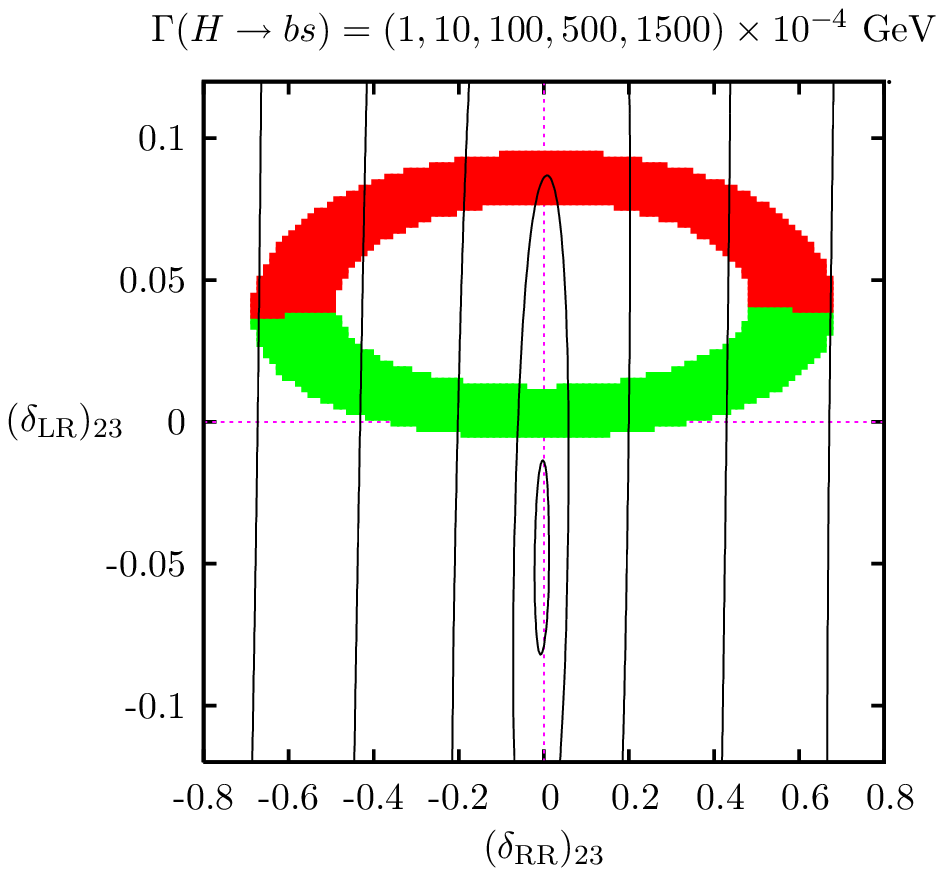} \\[1ex]
\includegraphics[height=.42\hsize]{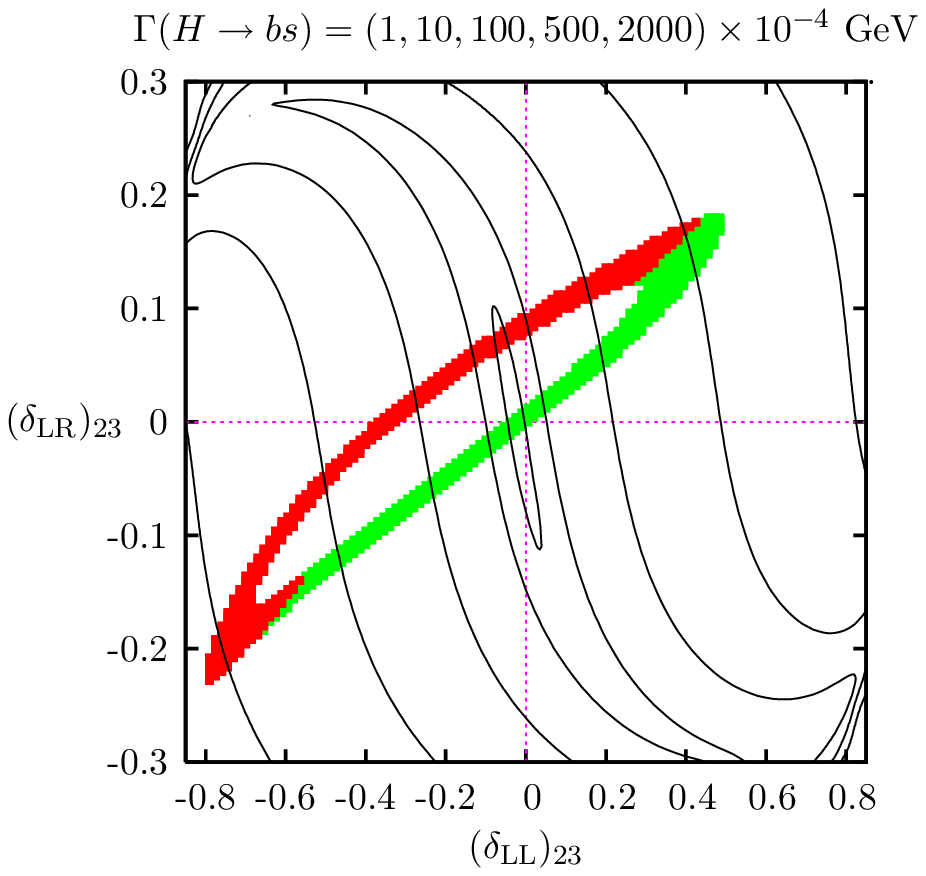} &
\includegraphics[height=.42\hsize]{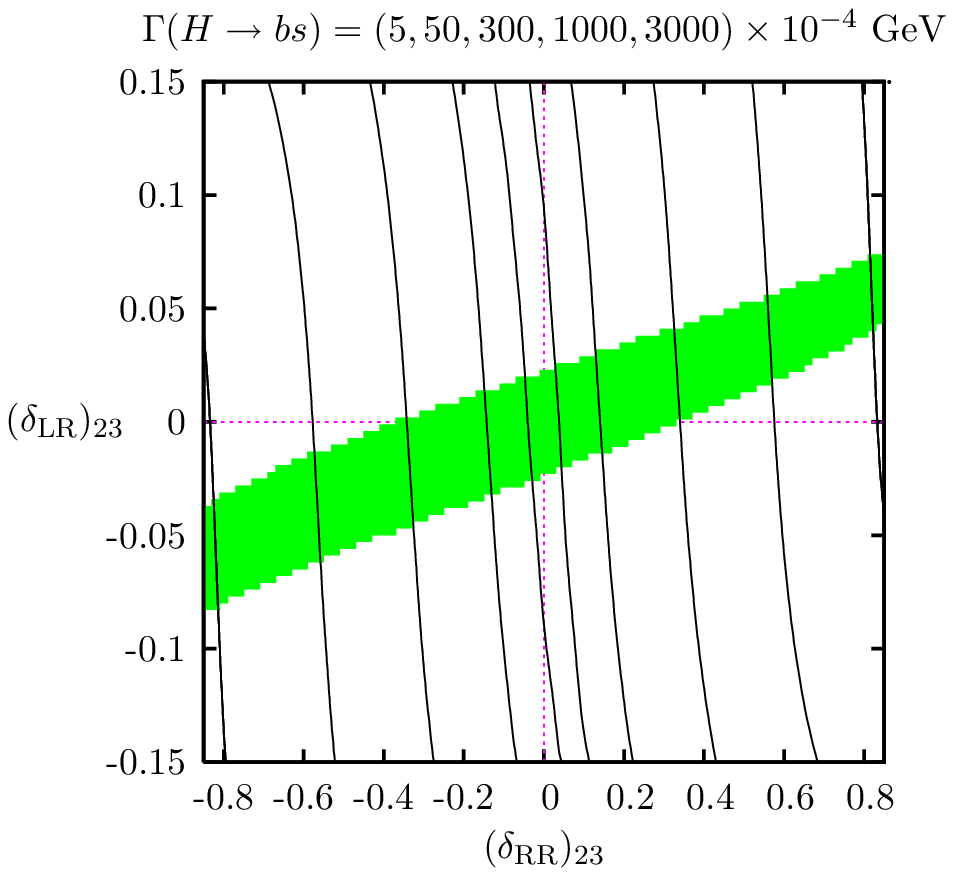} \\[1ex]
\includegraphics[height=.42\hsize]{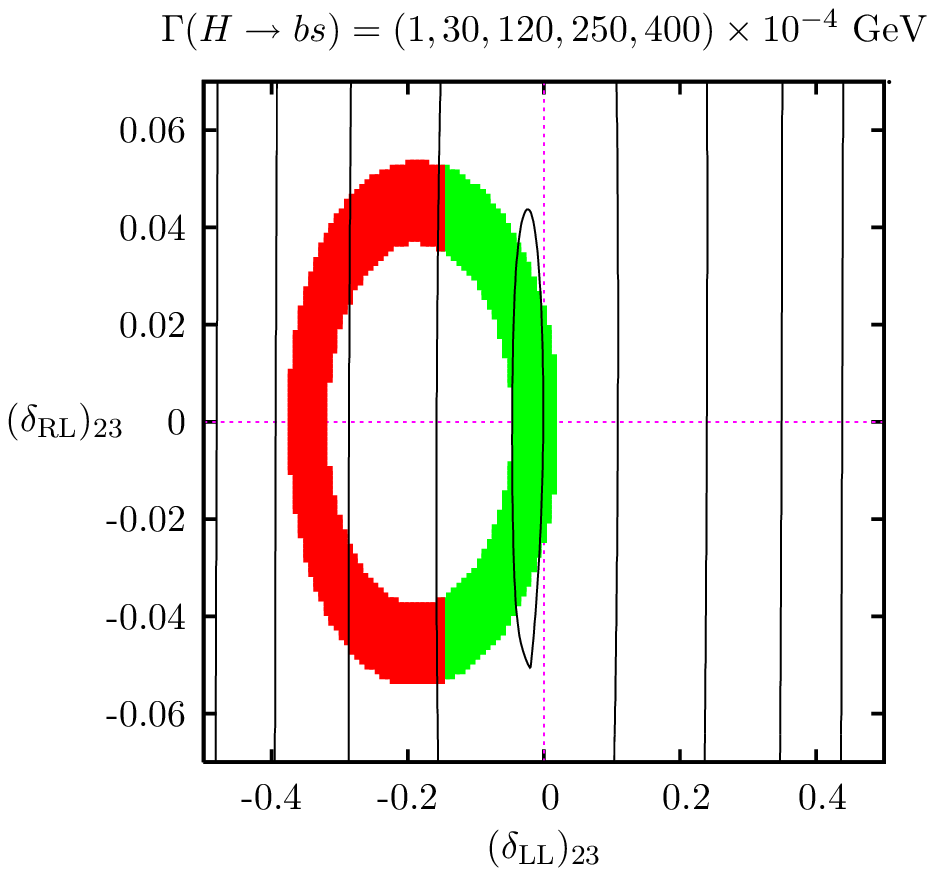} &
\includegraphics[height=.42\hsize]{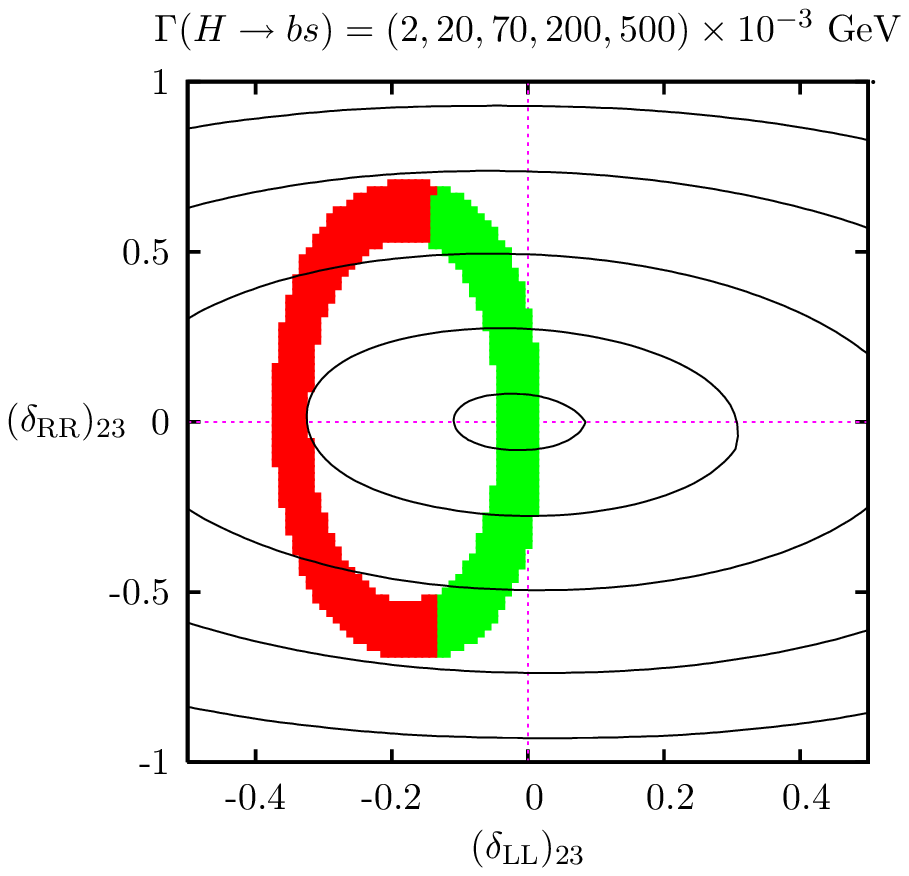}
\end{tabular}
\end{center}
\vspace*{-3ex}
\caption{\label{fig:planes}Contours of constant $\Gamma(\Hzbs)$ in 
various planes of the flavour-mixing parameters $\de{ab}$.  The coloured
bands indicate regions experimentally allowed by $B\to X_s\gamma$. The red
bands show regions disfavoured by $B\to X_s\mu^+\mu^-$.}
\end{figure}

Fig.~\ref{fig:planes} displays the results for our parameter set
(\ref{eq:numparameters}).  Drawn are contours of constant
$\Gamma(\Hzbs)\equiv \Gamma(H^0\to b\bar s) + \Gamma(H^0\to s\bar b)$
for various combinations $\de{ab}$--$\de{cd}$ of flavour-mixing
parameters, which we shall refer to as ``$ab$--$cd$ planes'' for short
in the following.  The coloured bands represent regions experimentally
allowed by $B\to X_s\gamma$. The red bands are regions disfavoured by
$B\to X_s\mu^+\mu^-$.

Contributions in the $\LR$--$\RL$ and $\LR$--$\RR$ planes lead to
maximal values
\begin{align}
\Gamma(\Hzbs)\sub{max} &= 3\times 10^{-4}\text{ GeV}
&\text{for } \de\LR &= 0.04, & \de\RL &= 0.04\,, \\
\Gamma(\Hzbs)\sub{max} &= 0.15\text{ GeV}
&\text{for } \de\LR &= 0.035, & \de\RR &= \pm0.7\,.
\end{align}
This translates to branching ratios compatible with experimental data 
of $\BR(\Hzbs)\sub{max}\sim 10^{-5}$ in the first, and 
$\Ord(10^{-3})$ in the second case.  Here we have used the total 
width of $\Gamma(H\to X)\approx 26$ GeV, $H = H^0, A^0$, for the point
(\ref{eq:numparameters}) in the MSSM with MFV.

The bounds on $\de\LR$, the best constrained for only one non-zero
flavour-off-diagonal element, are dramatically relaxed when
other flavour-changing parameters contribute simultaneously.  Values of
$\de\LR\sim 10^{-1}$ are allowed.  In particular, large although
fine-tuned values of $\de\LL$ and $\de\LR$ combined are not excluded by
$\bsg$, yielding e.g.
\begin{align}
\Gamma(\Hzbs)\sub{max} &= 0.25\text{ GeV}
&\text{for } \de\LR &= -0.22, & \de\LL &= -0.8\,, \\
\Gamma(\Hzbs)\sub{max} &= 0.35\text{ GeV}
&\text{for } \de\RL &= \pm0.04, & \de\RR &= -0.15\,.
\end{align}
The branching ratio is $\BR(\Hzbs)\sub{max}\sim 10^{-2}$ in both
cases.

Finally, we studied the combined effects of $\RL$--$\LL$ and
$\RR$--$\LL$.  We obtain
\begin{align}
\Gamma(\Hzbs)\sub{max} &= 2.5\times 10^{-3}\text{ GeV}
&\text{for } \de\RL &= 0.006, & \de\LL &= -0.128\,, \\
\Gamma(\Hzbs)\sub{max} &= 0.12\text{ GeV}
&\text{for } \de\RR &= 0.65, & \de\LL &= \pm0.14\,.
\end{align}
This means $\BR(\Hzbs)\sub{max}\sim 10^{-4}$ for the $\RL$--$\LL$
and $\BR(\Hzbs)\sub{max}\sim 10^{-2}$ for the $\RR$--$\LL$ case.

Therefore, we conclude that the predictions on $\BR(\Hzbs)$ induced by
$\de\RR$ or $\de\LL$ only, of $\Ord(10^{-3})$ or $\Ord(10^{-5})$, are greatly 
exceeded by the combination of $\de\LL$ and $\de\LR$ or $\de\RR$ and $\de\RL$ or
$\de\LL$ and $\de\RR$, which are of $\Ord(10^{-2})$.  Values of
$\BR(\Hzbs)\sim \Ord(10^{-3})$ emerge when considering the $\LR$--$\RR$.  
Thus, the destructive interference of the
combined set of flavour-mixing parameters leads to large allowed values
for the $\Hzbs$ branching ratio, which in general can be two orders of
magnitude larger than if induced by just one flavour-mixing parameter. 

\section{Conclusions}

In this work we carried out a phenomenological analysis of the
predictions for flavour-changing neutral heavy Higgs decays $H\to b\bar
s,s\bar b$, taking into account the constraints from $\bsg$ on the
flavour-mixing parameters $\de{\LL,\LR,\RL,\RR}$ of the MSSM with NMFV. 
We followed a fully diagrammatic approach at one loop, valid for
arbitrary values of $\tan\beta$ and the $\delta$'s, as long as they are
kinematically allowed.  The calculations were performed with
\textsl{FeynArts}, \textsl{FormCalc}, and \textsl{LoopTools}, which were
extended to include the MSSM with NMFV in a new model file and the
corresponding $6\times 6$ diagonalization, both of which are now
available in the public distributions of \textsl{FeynArts} and
\textsl{FormCalc}.  The contributions and interfering effects of the SM
and the various sectors of the MSSM were fully accounted for and the
interplay between the different types of non-minimal flavour mixing
explored, both individually and in combinations. 

We found that $\BR(\Hbs)$, whose value is at most of $\Ord(10^{-6})$
when induced by $\de\RL$ or $\de\LR$, can be of $\Ord(10^{-5})$ 
or $\Ord(10^{-3})$ if induced by $\de\RR$ or $\de\LL$.  By imposing
the constraints from $\bsg$ over the flavour-changing parameters, the
$\Hbs$ branching ratio depends only weakly on $\de\RL$ and $\de\LR$. 
The combined effect of two of the flavour-mixing parameters may raise
the predictions for $\BR(\Hbs)$ by one or two orders of magnitude,
typically $\Ord(10^{-3}$--$10^{-2})$.

Overall, the predictions for $H\to b\bar{s}$ are more optimistic when
the bounds from $\bsg$ are accounted for in the more realistic case of
several flavour-mixing parameters contributing simultaneously.


\section*{Acknowledgements}

We thank D.~St\"ockinger, S.~Heinemeyer, J.~Guasch and J.~Prades 
for very helpful discussions.  The work of S.P.\ has been supported by the 
European Union under contract No.~MEIF-CT-2003-500030.  J.I.I.\ acknowledges the
financial support by the E.U.\ (HPRN-CT-2000-149), the Spanish MCYT
(FPA2003-09298-C02-01) and Junta de Andaluc{\'\i}a (FQM-101).


\begin{appendix}

\section{NMFV Feynman Rules}
\label{sect:frules}

The following table lists Feynman rules needed for our computation in
the MSSM with NMFV.  These vertices were generated automatically from
the new \textsl{FeynArts} model file FVMSSM.mod, available from
www.feynarts.de.

The upper and lower entries of the FFS couplings are the prefactors of
the left- and right-handed chirality projectors.  The matrices
$T_{ab}^g$ are the SU(3) generators with gluon index $g$ and colour
indices $a$, $b$.  The matrices $U$, $V$ ($Z$) are the chargino
(neutralino) mixing matrices and CKM is the Cabibbo--Kobayashi--Maskawa
quark mixing matrix, with $i,j$ the fermion and $\rho,\sigma$ the squark
generation indices.  The angle $\alpha$ is the mixing angle in the 
neutral Higgs sector.

\begin{small}
\noindent
\begin{longtable}{p{.97\linewidth}}
\Coupling{FFS}{Gluino -- Quark -- Squark}
$
{C}(\tilde g,\Bar{u_i},\tilde u_{\sigma}) = { }
\begin{CoupVec}
\sqrt 2\,\ii\,g_s\,R_{\sigma,3 + {i}}^{\tilde u*}\,T_{ab}^g\\
\\[-3ex]
-\sqrt 2\,\ii\,g_s\,R_{\sigma,{i}}^{\tilde u*}\,T_{ab}^g
\end{CoupVec}
$\\
\bigskip
$
{C}(\tilde g,\Bar{d_i},\tilde d_{\sigma}) = { }
\begin{CoupVec}
\sqrt 2\,\ii\,g_s\,R_{\sigma,3 + {i}}^{\tilde d*}\,T_{ab}^g\\
\\[-3ex]
-\sqrt 2\,\ii\,g_s\,R_{\sigma,{i}}^{\tilde d*}\,T_{ab}^g
\end{CoupVec}
$\\
\bigskip
$
{C}(\tilde g,u_i,\tilde u_{\sigma}^{\dagger}) = { }
\begin{CoupVec}
-\sqrt 2\,\ii\,g_s\,R_{\sigma,{i}}^{\tilde u}\,T_{ba}^g\\
\\[-3ex]
\sqrt 2\,\ii\,g_s\,R_{\sigma,3 + {i}}^{\tilde u}\,T_{ba}^g
\end{CoupVec}
$\\
\bigskip
$
{C}(\tilde g,d_i,\tilde d_{\sigma}^{\dagger}) = { }
\begin{CoupVec}
-\sqrt 2\,\ii\,g_s\,R_{\sigma,{i}}^{\tilde d}\,T_{ba}^g\\
\\[-3ex]
\sqrt 2\,\ii\,g_s\,R_{\sigma,3 + {i}}^{\tilde d}\,T_{ba}^g
\end{CoupVec}
$\\
\bigskip
\Coupling{FFS}{Chargino -- Quark -- Squark}
$
{C}(\tilde \chi_{{c}}^{-},\Bar{d_j},\tilde u_{\sigma}) = { }
\begin{CoupVec}
\sum\limits_{{i}=1}^{3}
\frac{\ii\,e\,m_{d_j}\,{\rm CKM}_{ij}^{*}\,R_{\sigma,{i}}^{\tilde u*}\,U_{c2}^{*}}{\sqrt 2\,c_\beta\,M_W\,s_W}\\
\\[-3ex]
-\sum\limits_{{i}=1}^{3}
\frac{\ii\,e\,{\rm CKM}_{ij}^{*}}{2\,M_W\,s_\beta\,s_W}\,\left( 2\,M_W\,R_{\sigma,{i}}^{\tilde u*}\,s_\beta\,V_{c1} - \sqrt 2\,m_{u_i}\,R_{\sigma,3 + {i}}^{\tilde u*}\,V_{c2} \right) 
\end{CoupVec}
$\\
\bigskip
$
{C}(\tilde \chi_{{c}}^{+},\Bar{u_i},\tilde d_{\sigma}) = { }
\begin{CoupVec}
\sum\limits_{{j}=1}^{3}
\frac{\ii\,e\,m_{u_i}\,{\rm CKM}_{ij}\,R_{\sigma,{j}}^{\tilde d*}\,V_{c2}^{*}}{\sqrt 2\,M_W\,s_\beta\,s_W}\\
\\[-3ex]
-\sum\limits_{{j}=1}^{3}
\frac{\ii\,e\,{\rm CKM}_{ij}}{2\,c_\beta\,M_W\,s_W}\,\left( 2\,c_\beta\,M_W\,R_{\sigma,{j}}^{\tilde d*}\,U_{c1} - \sqrt 2\,m_{d_j}\,R_{\sigma,3 + {j}}^{\tilde d*}\,U_{c2} \right) 
\end{CoupVec}
$\\
\bigskip
$
{C}(d_j,\tilde \chi_{{c}}^{+},\tilde u_{\sigma}^{\dagger}) = { }
\begin{CoupVec}
-\sum\limits_{{i}=1}^{3}
\frac{\ii\,e\,{\rm CKM}_{ij}}{2\,M_W\,s_\beta\,s_W}\,\left( 2\,M_W\,R_{\sigma,{i}}^{\tilde u}\,s_\beta\,V_{c1}^{*} - \sqrt 2\,m_{u_i}\,R_{\sigma,3 + {i}}^{\tilde u}\,V_{c2}^{*} \right) \\
\\[-3ex]
\sum\limits_{{i}=1}^{3}
\frac{\ii\,e\,m_{d_j}\,{\rm CKM}_{ij}\,R_{\sigma,{i}}^{\tilde u}\,U_{c2}}{\sqrt 2\,c_\beta\,M_W\,s_W}
\end{CoupVec}
$\\
\bigskip
$
{C}(u_i,\tilde \chi_{{c}}^{-},\tilde d_{\sigma}^{\dagger}) = { }
\begin{CoupVec}
-\sum\limits_{{j}=1}^{3}
\frac{\ii\,e\,{\rm CKM}_{ij}^{*}}{2\,c_\beta\,M_W\,s_W}\,\left( 2\,c_\beta\,M_W\,R_{\sigma,{j}}^{\tilde d}\,U_{c1}^{*} - \sqrt 2\,m_{d_j}\,R_{\sigma,3 + {j}}^{\tilde d}\,U_{c2}^{*} \right) \\
\\[-3ex]
\sum\limits_{{j}=1}^{3}
\frac{\ii\,e\, m_{u_i}\,{\rm CKM}_{ij}^{*}\,R_{\sigma,{j}}^{\tilde d}\,V_{c2}}{\sqrt 2\,M_W\,s_\beta\,s_W}
\end{CoupVec}
$\\
\bigskip
\Coupling{FFS}{Neutralino -- Quark -- Squark}
$
{C}(\tilde \chi_{{n}}^{0},\Bar{u_i},\tilde u_{\sigma}) = \newline
\begin{CoupVec}
\frac{\ii\,e}{3\,\sqrt 2\,c_W\,M_W\,s_\beta\,s_W}\,
\left( 4\,M_W\,R_{\sigma,3 + {i}}^{\tilde u*}\,s_\beta\,s_W\,Z_{n1}^{*} - 3\,c_W\,m_{u_i}\,R_{\sigma,{i}}^{\tilde u*}\,Z_{n4}^{*} \right) \\
\\[-3ex]
-\frac{\ii\,e}{3\,\sqrt 2\,c_W\,M_W\,s_\beta\,s_W}\,
\left( 3\,c_W\,m_{u_i}\,R_{\sigma,3 + {i}}^{\tilde u*}\,Z_{n4} + M_W\,R_{\sigma,{i}}^{\tilde u*}\,s_\beta\,\left( s_W\,Z_{n1} + 3\,c_W\,Z_{n2} \right)  \right) 
\end{CoupVec}
$\\
\bigskip
$
\vrule width 0ex height 6ex depth 0ex
{C}(\tilde \chi_{{n}}^{0},\Bar{d_i},\tilde d_{\sigma}) = \newline
\begin{CoupVec}
-\frac{\ii\,e}{3\,\sqrt 2\,c_\beta\,c_W\,M_W\,s_W}\,
\left( 2\,c_\beta\,M_W\,R_{\sigma,3 + {i}}^{\tilde d*}\,s_W\,Z_{n1}^{*} + 3\,c_W\,m_{d_i}\,R_{\sigma,{i}}^{\tilde d*}\,Z_{n3}^{*} \right) \\
\\[-3ex]
-\frac{\ii\,e}{3\,\sqrt 2\,c_\beta\,c_W\,M_W\,s_W}\,
\left( 3\,c_W\,m_{d_i}\,R_{\sigma,3 + {i}}^{\tilde d*}\,Z_{n3} + c_\beta\,M_W\,R_{\sigma,{i}}^{\tilde d*}\,\left( s_W\,Z_{n1} - 3\,c_W\,Z_{n2} \right)  \right) 
\end{CoupVec}
$\\
\bigskip
$
\vrule width 0ex height 6ex depth 0ex
{C}(u_i,\tilde \chi_{{n}}^{0},\tilde u_{\sigma}^{\dagger}) = \newline
\begin{CoupVec}
-\frac{\ii\,e}{3\,\sqrt 2\,c_W\,M_W\,s_\beta\,s_W}\,
\left( M_W\,R_{\sigma,{i}}^{\tilde u}\,s_\beta\,s_W\,Z_{n1}^{*} + 3\,c_W\,\left( M_W\,R_{\sigma,{i}}^{\tilde u}\,s_\beta\,Z_{n2}^{*} + m_{u_i}\,R_{\sigma,3 + {i}}^{\tilde u}\,Z_{n4}^{*} \right)  \right) \\
\\[-3ex]
\frac{\ii\,e}{3\,\sqrt 2\,c_W\,M_W\,s_\beta\,s_W}\,
\left( 4\,M_W\,R_{\sigma,3 + {i}}^{\tilde u}\,s_\beta\,s_W\,Z_{n1} - 3\,c_W\,m_{u_i}\,R_{\sigma,{i}}^{\tilde u}\,Z_{n4} \right) 
\end{CoupVec}
$\\
\bigskip
$
\vrule width 0ex height 6ex depth 0ex
{C}(d_i,\tilde \chi_{{n}}^{0},\tilde d_{\sigma}^{\dagger}) = \newline
\begin{CoupVec}
-\frac{\ii\,e}{3\,\sqrt 2\,c_\beta\,c_W\,M_W\,s_W}\,
\left( c_\beta\,M_W\,R_{\sigma,{i}}^{\tilde d}\,s_W\,Z_{n1}^{*} - 3\,c_W\,\left( c_\beta\,M_W\,R_{\sigma,{i}}^{\tilde d}\,Z_{n2}^{*} - m_{d_i}\,R_{\sigma,3 + {i}}^{\tilde d}\,Z_{n3}^{*} \right)  \right) \\
\\[-3ex]
-\frac{\ii\,e}{3\,\sqrt 2\,c_\beta\,c_W\,M_W\,s_W}\,
\left( 2\,c_\beta\,M_W\,R_{\sigma,3 + {i}}^{\tilde d}\,s_W\,Z_{n1} + 3\,c_W\,m_{d_i}\,R_{\sigma,{i}}^{\tilde d}\,Z_{n3} \right) 
\end{CoupVec}
$\\
\bigskip
\Coupling{SSS}{Higgs -- 2 Squarks}
$
{C}(H^{0},\tilde u_{\rho},\tilde u_{\sigma}^{\dagger}) =\newline
{-}\displaystyle\sum\limits_{{i},{j}=1}^{3}\frac{\ii\,e}{6\,c_W\,M_W\,s_\beta\,s_W}\, 
\begin{PlusB}
R_{\rho,3 + {i}}^{\tilde u*}\, 
\begin{PlusB}
4\,c_{\alpha+\beta}\,\delta_{ij}\,M_W\,M_Z\,R_{\sigma,3 + {j}}^{\tilde u}\,s_\beta\,s_W^{2} \\
- 3\,c_W\,R_{\sigma,{j}}^{\tilde u}\,( c_{\alpha}\,\delta_{ij}\,m_{u_i}\,\mu - m_{u_j}\,s_{\alpha}\,A^{u*}_{ji} ) \,\\
+6\,c_W\,\delta_{ij}\,m_{u_i}^{2}\,R_{\sigma,3 + {j}}^{\tilde u}\,s_{\alpha}
\end{PlusB}\,\\
-R_{\rho,{i}}^{\tilde u*}\,
\begin{PlusB} 3\,c_W\,R_{\sigma,3 + {j}}^{\tilde u}\,( c_{\alpha}\,\delta_{ij}\,m_{u_i}\,\mu^{*} - A^u_{ij}
\,m_{u_i}\,s_{\alpha})  \\
 - \delta_{ij}\,R_{\sigma,{j}}^{\tilde u}\,( 6\,c_W\,m_{u_i}^{2}\,s_{\alpha} + c_{\alpha+\beta}\,M_W\,M_Z\,s_\beta\,( 3 - 4\,s_W^{2} ) ) \end{PlusB}
\end{PlusB}
$\\
\bigskip
$
\vrule width 0ex height 6ex depth 0ex
{C}(H^{0},\tilde d_{\rho},\tilde d_{\sigma}^{\dagger}) = \newline
\displaystyle\sum\limits_{{i},{j}=1}^{3}
\frac{{\ii}\,e}{6\,c_\beta\,c_W\,M_W\,s_W}\,{ }
\begin{PlusB}
R_{\rho,3 + {i}}^{\tilde d*}\, 
\begin{PlusB}
2\,c_{\alpha+\beta}\,c_\beta\,\delta_{ij}\,M_W\,M_Z\,R_{\sigma,3 + {j}}^{\tilde d}\,s_W^{2} \\
+ 3\,c_W\,R_{\sigma,{j}}^{\tilde d}\,( \delta_{ij}\,m_{d_i}\,\mu\,s_{\alpha} - c_{\alpha}\,m_{d_j}\, A^{d*}_{ji}) \,\\
-6\,c_{\alpha}\,c_W\,\delta_{ij}\,m_{d_i}^{2}\,R_{\sigma,3 + {j}}^{\tilde d}
\end{PlusB}\,\\
-R_{\rho,{i}}^{\tilde d*}\,
\begin{PlusB}
 3\,c_W\,R_{\sigma,3 + {j}}^{\tilde d}\,( A^d_{ij}\,c_{\alpha}\,m_{d_i} - \delta_{ij}\,m_{d_i}\,\mu^{*}\,s_{\alpha} )  \\
+ \delta_{ij}\,R_{\sigma,{j}}^{\tilde d}\,( 6\,c_{\alpha}\,c_W\,m_{d_i}^{2} - c_{\alpha+\beta}\,c_\beta\,M_W\,M_Z\,( 3 - 2\,s_W^{2}))
\end{PlusB} \end{PlusB}
$\\
\bigskip
\end{longtable}
\end{small}

\end{appendix}


\end{document}